# Estimation of the incidence rate and mortality rate ratio for chronic conditions based on aggregated current status data

Ralph Brinks

Chair for Medical Biometry and Epidemiology, Witten/Herdecke University, Faculty of Health/School of Medicine, D-58448 Witten, Germany

Correspondence:
E-Mail: ralph.brinks@uni-wh.de

## Abstract

Recently, it has been shown that the transition rates of the illness-death model (IDM) for chronic conditions are related to the percentages of people in the states by a three-dimensional system of differential equations [Bri24]. The aim of this article is to introduce a method to estimate the age-specific incidence rate together with the mortality rate ratio from aggregated current status (ACS) data. By ACS data we mean counts of (non-necessarily different) people in the three states of the IDM at different points in time. ACS data stem from epidemiological studies where only current disease status and vital status data need to be collected without following-up people (as, for example, in cohort studies). As an application, we use the theory in a simulation study about diabetes in Germany with 600 study subjects at eleven repeated cross-sections each of which with 50% participation quote. Special focus is given to stochastic dependency of the sampled participants. We find a good agreement between the estimates and the input parameters used for the simulation.

Key words: illness-death model, prevalence, chronic diseases, diabetes, epidemiology, study design, bootstrapping. differential equations.

# Introduction

It could be shown recently that the transition rates in the illness-death model (IDM) for chronic diseases (see Figure 1) are linked to the percentages of people in the three states *Non-diseased*, *Diseased* and *Dead* via a linear three-dimensional system of ordinary differential equations (ODEs) [Bri24]. The derivation of the ODE system uses the theory of the chemical master equation for stochastic modeling of biochemical reaction systems [Jah07].

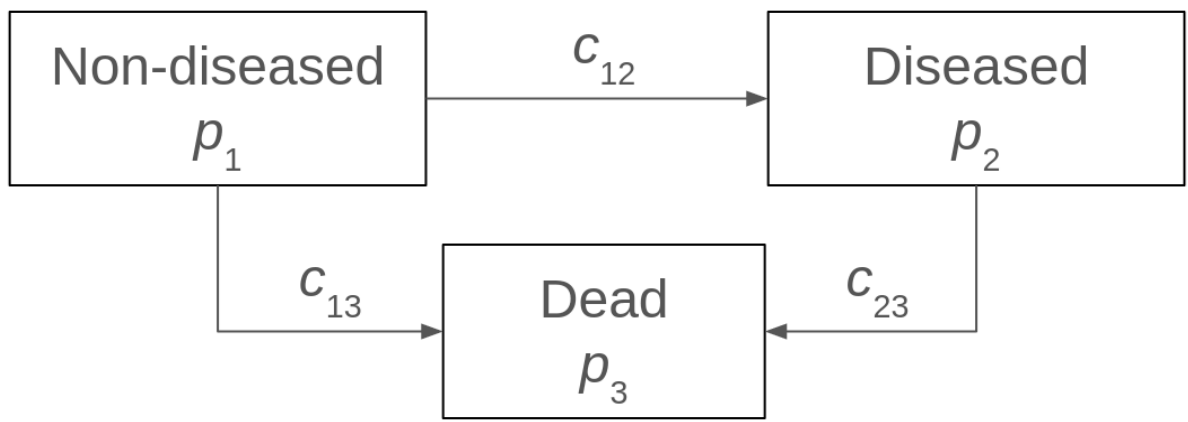

**Figure 1: Illness-death model for chronic diseases with transition rates $c_{jk}$ and fractions $p_j$ of people in the three states.**

The aim of this article is applying the theory of the previous article [Bri24] to a problem from epidemiology: in order to obtain information about the transition rates in the IDM, usually follow-up studies are run. A group of initially disease-free people (*Non-diseased* in Figure 1) is included as participants in a study and is followed over time. At predetermined points in time, the participants are invited and examined whether the considered disease has evolved. If a study participant has contracted the disease between two consecutive examinations, the subject is counted as an incident case. Appropriate methods for statistical analysis in this situation can be found, for example, in [Hou16]. Collecting this type of data requires follow-up of the study participants over time, which sometimes can be quite costly due to administrative and logistic efforts and, in case of rare diseases, also rather lengthy (because incident cases are scarce). In contrast to follow-up studies, cross-sectional studies examine disease and vital status of people at predetermined points in time. Here we assume that at each point in time, the number of people in one of the states of the IDM (*Non-diseased*, *Diseased*, *Dead*) is surveyed. The collected data of each study participant is the current status at the specific point in time; for a diseased subject it is irrelevant when the disease became prevalent. For each point in time, we obtain count data for each of the three states, which explains the name aggregated current status (ACS) data. An example for ACS data is shown in Table 1, which reads, for example, that at time $t^* = 80$ (years) 184 subjects are deceased.

| **State** | **Time *t*** | | | | | | | | | | |
|---|---|---|---|---|---|---|---|---|---|---|---|
| | **0** | **10** | **20** | **30** | **40** | **50** | **60** | **70** | **80** | **90** | **100** |
| Non-diseased | 325 | 285 | 300 | 291 | 275 | 262 | 233 | 155 | 68 | 16 | 0 |
| Diseased | 0 | 0 | 0 | 0 | 7 | 15 | 43 | 63 | 41 | 8 | 0 |
| Dead | 0 | 0 | 1 | 1 | 4 | 8 | 27 | 81 | 184 | 260 | 298 |
| ∑ | 325 | 285 | 391 | 292 | 286 | 285 | 303 | 299 | 293 | 284 | 298 |

**Table 1: Example for aggregated current status (ACS) data from the IDM at eleven different points in time *t* = 0, …, 100.**

Note that the individual subjects at the different points in time *t* are not assumed to be all the same or all different. To increase the stochastic dependency in the data set, a complex visit schema has been simulated, which is detailed in the section "Simulation" below. As a result, not even the total numbers of participants at the time points (last row of Table 1) are the same, they range from 284 (at *t* = 90) to 325 (at *t* = 0).

## Illness-death model and the system of ODEs

We consider the IDM as depicted in Figure 1. Each subject of the population under consideration is assigned to exactly one health state: *Non-diseased*, *Diseased* and *Dead*. Subjects may change their state along the arrows in Figure 1 as time *t* evolves. The non-negative transition rates are denoted by $c_{jk}$ and usually depend on time *t*, i.e., $c_{jk} = c_{jk}(t)$. The fractions of people in the three states are denoted by $p_j = p_j(t)$, *j* = 1, 2, 3. For example, at some time *t*\* the fraction $p_3(t^*)$ denotes the fraction of people, who are deceased at *t*\*. The fractions $p_1, p_2,$ and $p_3$ always add up to 100%: $p_1(t) + p_2(t) + p_3(t) = 1$ for all *t*.

The most important finding of [Jah07] for the IDM states that if the initial condition at, say *t* = 0, is the multinomial distribution $\mathbb{P}(0, x) = \mathcal{M}(x, N, p_0)$ for a parameter vector $p_0 \in [0,1]^3$, then for *t* > 0 it holds $\mathbb{P}(t, x) = \mathcal{M}(x, N, p(t))$ where the parameter $p(t)$ is the solution of the following system of ordinary differential equations (ODEs):

$$\frac{d}{dt} p(t) \;=\; p'(t) \;=\; A(t)\, p(t) \qquad (1)$$

with initial condition $p(0) = p_0$. The matrix $A(t)$ in Eq. (1) is given by

$$A(t) \;=\; \begin{bmatrix} -c_{12}(t) - c_{13}(t) & 0 & 0 \\ c_{12}(t) & -c_{23}(t) & 0 \\ c_{13}(t) & c_{23}(t) & 0 \end{bmatrix}$$

Using $p(t) = (p_1(t), p_2(t), p_3(t))$, we obtain following linear system of ODEs:

$$p_1' = -(c_{12} + c_{13})\, p_1 \qquad (2a)$$
$$p_2' = c_{12}\, p_1 - c_{23}\, p_2 \qquad (2b)$$
$$p_3' = c_{13}\, p_1 + c_{23}\, p_2. \qquad (2c)$$

Note that the initial condition $p_0 \in [0,1]^3$ is three-dimensional, but $p_1, p_2, p_3 \in [0,1]$ are scalars.

## Simulation

To apply the theory presented in the previous section, we choose a test example from [Bri18] motivated by the situation of type 2 diabetes in Germany. Diabetes is assumed to be irreversible. For setting up the simulation, the incidence rate $c_{12}$ is given by $c_{12}(t) = \max(0, t - 30)/2000$ and the mortality rates $c_{13}$ and $c_{23}$ are assumed to be of Gompertz-type $c_{13}(t) = \exp(-10.7 + 0.1\, t)$ and $c_{23}(t) = \exp(-10 + 0.1\, t)$.

We mimic a study with about 300 participants surveyed at time points $t_1 = 0$, $t_2 = 10$, ..., $t_{11} = 100$. Usually, potential participants of a study are registered and invited for a visit for a medical examination. Not all invited participants join the examination visits at each of the time points $t_k$, $k = 1, ..., K$. Here a participation probability of $p_{part} = 50\%$ at each time $t_k$, $k = 1, ..., K$, for a total of $N = 600$ invited subjects is assumed: for each of the $N = 600$ potential participants, a 1:1 chance is simulated if the participant joins the examination visit at $t_k$ or not.

The simulation starts with the $N = 600$ potential study participants, for whom the changes of the states in the IDM as in Figure 1 is simulated by the method described in [Bri14]. After this, for each of the $N = 600$ potential study participants and each of the $K$ points in time, we simulate whether the potential participant joins the examination visit according to $p_{part}$. Table 2 shows how frequent the $N = 600$ potential participants join the eleven examination visits. For example, none of the 600 potential participants joined all examinations, and 137 joined 5 of the 11 examinations.

| Number of visits | 0 | 1 | 2 | 3 | 4 | 5 | 6 | 7 | 8 | 9 | 10 | 11 | Σ |
|---|---|---|---|---|---|---|---|---|---|---|---|---|---|
| Number of subjects | 1 | 5 | 24 | 49 | 94 | 137 | 123 | 96 | 60 | 10 | 1 | 0 | 600 |

**Table 2: Number of visits of $N = 600$ potential study participants at eleven different points in time.**

As a result, we obtain the count data as shown in Table 1, which are not independent, because $600 - 1 - 5 = 594$ of the $N = 600$ were sampled multiple times. The resulting fractions $p^{(obs)}$ are shown in Figure 2 as filled dots. For comparison, the associated components of the solution $p$ of the ODE system (1) are shown as lines.

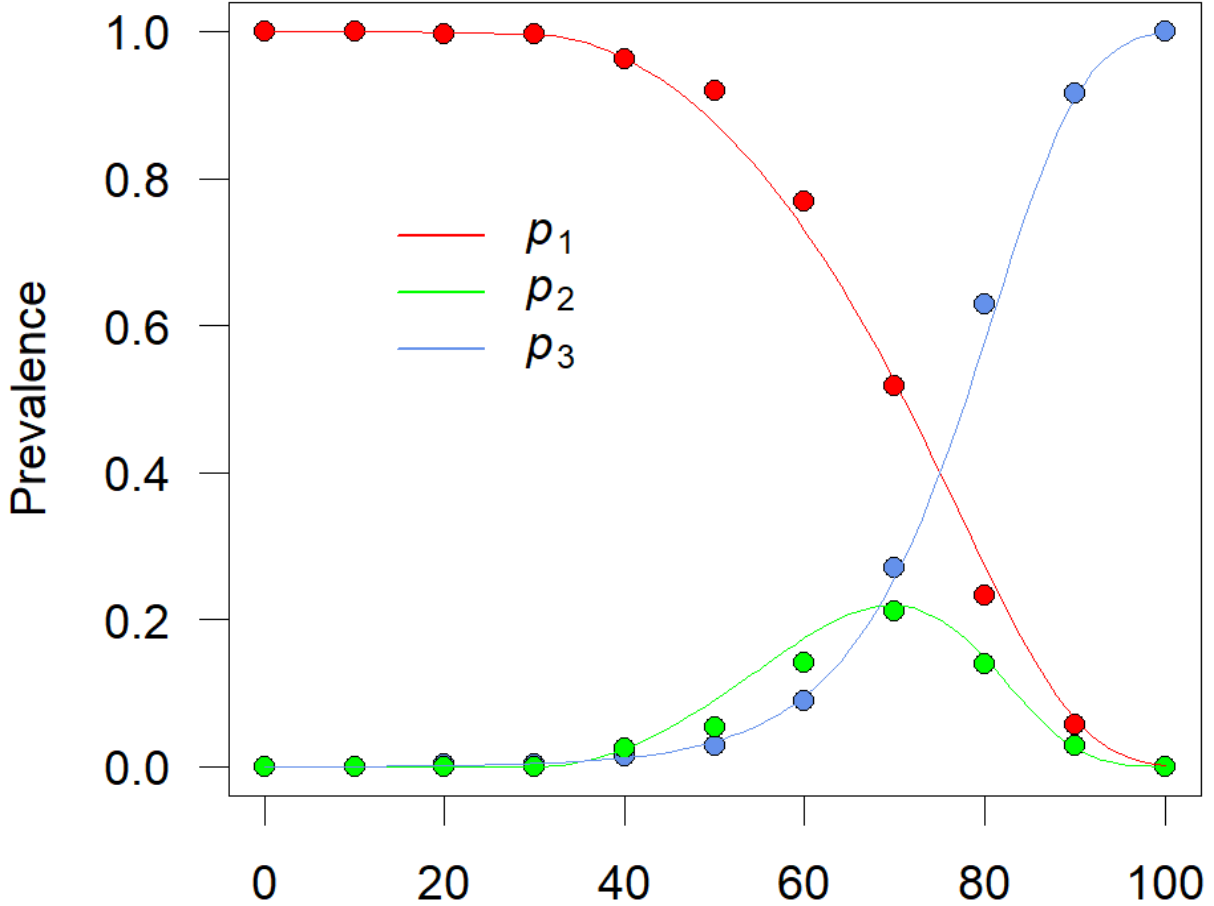

Figure 2: The three components of $p^{(obs)}$ from the ACS data in Table 1 are shown as filled dots. For comparison, the associated components of the solution $p$ of Equation (1) are shown as lines. The colors refer to the components, for example, green presents the second of the three components.

## Estimation

We want to use the system given in Eqs. (2a-c) to determine the incidence rate $c_{12}(t)$ and the mortality rate ratio (MRR) $c_{23}/c_{13}$ from ACS data as shown in Table 1 and graphically depicted as dots in Figure 2. As Eqs. (2a-c) are not independent ($p_3 = 1 - p_1 - p_2$), we obtain two independent equations and three unknowns $c_{12}$, $c_{13}$, and $c_{23}$. Hence, one additional piece of information is needed to make the system identifiable: we assume that the mortality rate $c_{13}$ is known and given by $c_{13}(t) = \exp(-10.7 + 0.1\ t)$. Under these circumstances, it is possible to formulate an estimation method for the incidence rate and the MRR. For this, we make a parametric approach with an unknown parameter $\vartheta = (\vartheta_1, \vartheta_2, \vartheta_3)$ by $c_{12}(t) = \vartheta_2 \times \max(0, t - \vartheta_1)$ and $c_{23}(t) = \vartheta_3 \times c_{13}(t)$. The parameter $\vartheta_1$ is the age of onset of diabetes, which for type 2 diabetes in Germany is at the age of about 30 years. The parameter $\vartheta_2$ is the slope of the age course of the incidence rate. As a coarse approximation of the age-specific incidence rate in Germany, we choose an incidence rate that increases linearly with age starting at age 30. The parameter $\vartheta_3$ is the MRR. The true parameter vector $\vartheta^{(true)}$ underlying the simulation in the previous section equals $\vartheta^{(true)} = (30, 1/2000, \exp(0.7))$. With other words, the solid lines in Figure 2 refer to the solution $p(t; \vartheta)$ of Equation (1) with initial condition $p(0) = (1, 0, 0)$ and $\vartheta^{(true)} = (30, 1/2000, \exp(0.7))$.

We consider two estimation methods for $\vartheta = (\vartheta_1, \vartheta_2, \vartheta_3)$ based on the data shown in Table 1, the first is based on a least squares fit, the second is a maximum likelihood approach.

## Least squares estimation

For a given $\vartheta$, we solve system (2a-c) and obtain the solution $p(t; \vartheta)$. Then, we calculate the squared residual $\| p(\ ; \vartheta) - p^{(obs)} \|^2$ of the difference between $p(\ ; \vartheta)$ and the observed fractions $p^{(obs)}$ from the ACS data. The norm $\| . \|^2$ sums over all observed points in time $t_k$, $k = 1, \ldots, K$, for example in Table 1 from $t_1 = 0$ to $t_{11} = 100$ in units of 10. The optimal $\vartheta^*$ is given by

$$\vartheta^* = \arg \min \| p(\ ; \vartheta) - p^{(obs)} \|^2. \tag{3}$$

In other words, $\vartheta^*$ minimizes the squared difference between the modelled solution $p(t; \vartheta)$ of system (2a-c) and the observed fractions $p^{(obs)} \in [0,1]^3$; hence, $\vartheta^*$ is a least squares estimator.

## Maximum likelihood estimation

In case that study participants at the $K$ cross-sections are all independent, the log-likelihood function $\ell(\vartheta)$ of the multinomial distribution can be used to estimate the unknown parameter $\vartheta = (\vartheta_1, \vartheta_2, \vartheta_3)$. Under independence we obtain

$$\ell(\vartheta) = \sum_{k=1}^{K} \left[ \log(n_k!) + \sum_{j=1}^{3} \left\{ x_{k,j} \log(p_j(t_k; \vartheta)) - \log(x_{k,j}!) \right\} \right], \tag{4}$$

which can be maximized with respect to $\vartheta$. In Eq. (4) $x_{k,j}$ is the number of subjects in state $j$, $j = 1, 2, 3$, at time $t_k$, $k = 1, \ldots, K$.

With a view to inference statistics, the question arises how confidence bounds for the parameter vector $\vartheta \in \mathbb{R}^3$ can be obtained. If the samples of the ACS data at the time points $t_k$, $k = 1, \ldots, K$, are stochastically independent (for example by having chosen random sampling), the likelihood function of the multinomial distribution $\mathcal{M}(x, N, p(t))$ can easily be formulated a the product of the $K$ probability mass functions similar to Eq. (4). Inversion of the Fisher information matrix may be used to calculate (asymptotic) confidence bounds. In general, however, we cannot assume independence of the samples at the time points $t_k$, $k = 1, \ldots, K$. Participants may possibly be re-examined at a later point in time and a deceased participant, of course, will be dead at all later points in time. Thus, we may have dependencies from one time point to another. For the general case, we suggest the following bootstrapping approach: Based on the parameter estimate $\vartheta^*$, we run a microsimulation of a population moving through the IDM given by Figure 1 with the transition rates $c_{12}$, $c_{13}$, and $c_{23}$ that come from the estimate $\vartheta^*$. This can be done by the algorithm described in [Bri14] or, for example, by the related Doob-Gillespie algorithm. Then, for each bootstrap we repeatedly mimic the exact sampling schema that has led to the ACS data. The visit schema described above is used for this purpose (see Table 2). If, for example, the ACS data at the different $t_k$, $k = 1, \ldots, K$, come from the same group of people, for each bootstrap we sample the same group of people. By this, we are able to model the same dependency in each bootstrap as the inherent stochastic dependency of the original ACS data.

For each bootstrap (indexed by $b$), we derive the three dimensional vector of fractions $p^{(\text{obs}, b)} \in [0,1]^3$ and estimate the least squares estimate $\vartheta^{\text{LS},b}$ according to Eq. (3) as well as the maximum likelihood estimator $\vartheta^{\text{ML},b}$ according to Eq. (4). Finally, we get a bootstrap population $\vartheta^{*,b}$, $b = 1, \ldots, B$, which can then be used to estimate the respective confidence bounds by the empirical quantile functions [Efr94].

For estimating $\vartheta$, we use the least-squares minimization as in Eq. (3) and maximum likelihood estimation as in Eq. (4). For a given $\vartheta$ the solution $p(t; \vartheta)$ in (3) and (4), is obtained by the classical Runge-Kutta method of fourth order (`rk4` in the R package `deSolve` [Soe10]). Estimation of 95% confidence bounds was done by the bootstrap algorithm described above with $B = 1000$ bootstraps. Figure 3 shows the resulting histograms for the three parameters $\vartheta_1$, $\vartheta_2$, and $\vartheta_3$ for the least-squares estimation (top row) and maximum likelihood estimation (bottom). The associated median together with the 2.5 and 97.5 quantiles are given in Table 3 for both estimation methods.

| Parameter | True value | Median and (2.5, 97.5)% Quantile | |
|---|---|---|---|
| | | **LS estimation** | **ML estimation** |
| $\vartheta_1$ | 30 | 33.8 (28.8, 38.4) | 33.7 (30.0, 37.3) |
| $\vartheta_2$ (per 10,000) | 5 | 6.29 (4.40, 9.09) | 6.30 (4.70, 8.40) |
| $\vartheta_3$ | 2.01 | 2.24 (1.71, 2.99) | 2.22 (1.79, 2.83) |

**Table 3: Median and quantiles of the components of $\vartheta$ in the $B = 1000$ bootstraps for the least-squares (LS) and the maximum-likelihood (ML) estimates.**

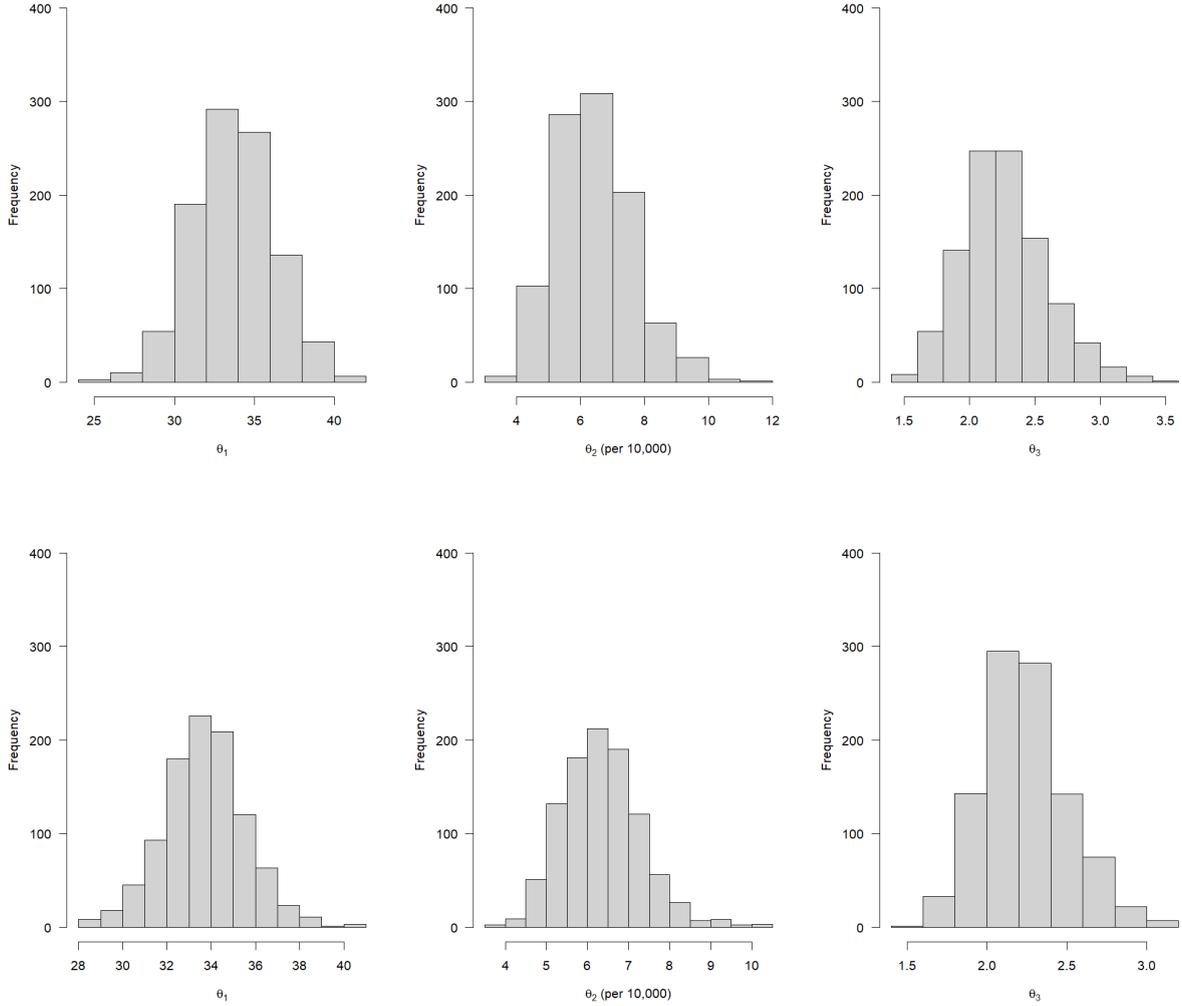

**Figure 3: Histograms of the three components of the $B = 1000$ bootstrap estimates $\vartheta^{*,b}$ in the least-squares (top row) and the maximum-likelihood estimation (bottom row).**

# Discussion

In this article, we have sketched how aggregated current status (ACS) data from the illness-death model (IDM) can be utilized to obtain information about the transition rates in the IDM. To estimate transition rates, usually cohort studies are run. ACS data, however, allude to a study design, where following-up study participants is not necessary. Compared to simple prevalence data, the number of deceased persons are necessary, which may be obtained from official residents' registries (in case they exist). Note that in epidemiology we frequently have the prevalence $\pi(t) := p_2(t)/[p_1(t) + p_2(t)]$ being reported in population studies. Inserting Eqs. (2a) and (2b) yields the ODE $\pi' = (1 - \pi)(c_{12} - \pi (c_{23} - c_{13})]$, which has the advantage of being scalar. Moreover, the incidence ($c_{12}$) can be expressed directly in terms of $\pi$, its derivative $\pi'$ and the excess mortality ($c_{23} - c_{13}$) [Bri18].

Apart from the two estimation techniques, we introduced a bootstrapping method for inference in the general case where samples are not stochastically independent. In our example, we calculated the 95% confidence intervals by empirical quantiles. Instead of empirical quantiles, other methods like the BCa method is possible [Efr94].

The question arises, where ACS data play a role in practical applications. ACS data has the advantage that in many cases individual study subjects are non-identifiable, because data are aggregated and need not be reported on the individual level. This can be important for data protection reasons - especially if the health state *Ill* refers to particularly sensible diseases like sexually transmitted diseases. As an example for an application of ACS data one might think of people with permanent need for long-term care in elderly population. In Germany, the mortality rate ratio (MRR) for these people is unknown. Follow-up studies are difficult because elderly people can be difficult to contact, withdrawal of consent is frequent and transport to examination centers imposes more problems than in a younger study population.

## Summary

We could show that the recently developed theory of [Bri24] can be used to obtain insights into the transition rates of illness-death model. Two methods for estimating the parameters have been developed, one method is based on a least-squares minimization and the other is based on maximum-likelihood optimization. Applicability has been demonstrated in a simulation study motivated by diabetes in Germany. We find a good agreement between both estimation methods and the input used to set up the simulations.